\documentclass[aps,prb,twocolumn,showpacs]{revtex4-2}
\usepackage{amssymb}
\usepackage{graphicx}
\usepackage{dcolumn}
\usepackage{bm}
\usepackage{amsmath}
\usepackage{soul,color}
\usepackage{textcomp}
\usepackage[table,xcdraw]{xcolor}
\usepackage{times}
\usepackage{tensor}
\definecolor{cGreen}{RGB}{0,0,0}
\definecolor{cBlue}{RGB}{45,51,180}
\definecolor{cmagenta}{RGB}{205,0,100}

\usepackage[colorlinks,linkcolor=blue,citecolor=blue]{hyperref}

\begin{document}

\title{Non-thermal dynamics in a spin-1/2 lattice Schwinger model}
\author{Chunping Gao$^{1}$}
\author{Zheng Tang$^{1}$}
\author{Fei Zhu$^{1}$}
\author{Yunbo Zhang$^{2}$}
\email{ybzhang@zstu.edu.cn}
\author{Han Pu$^{3}$}
\email{hpu@rice.edu}
\author{Li Chen$^{1}$}
\email{lchen@sxu.edu.cn}

\affiliation{
$^1${Institute of Theoretical Physics, State Key Laboratory of Quantum Optics and Quantum Optics Devices, Shanxi University, Taiyuan 030006, China}\\
$^2${Key Laboratory of Optical Field Manipulation of Zhejiang Province and Physics Department of Zhejiang Sci-Tech University, Hangzhou 310018, China}\\
$^3${Department of Physics and Astronomy, and Rice Center for Quantum Materials, Rice University, Houston, TX 77005, USA}
}

\begin{abstract}
Local gauge symmetry is intriguing for the study of quantum thermalization breaking. For example, in the high-spin lattice Schwinger model (LSM), the local U(1) gauge symmetry underlies the disorder-free many-body localization (MBL) dynamics of matter fields. This mechanism, however, would not work in a spin-1/2 LSM due to the absence of electric energy in the Hamiltonian. In this paper, we show that the spin-1/2 LSM can also exhibit disorder-free MBL dynamics, as well as entropy prethermalization, by introducing a four-fermion interaction into the system. The interplay between the fermion interaction and U(1) gauge symmetry endows the gauge fields with an effectively disordered potential which is responsible for the thermalization breaking. It induces anomalous (i.e., non-thermal) behaviors in the long-time evolution of such quantities as local observables, entanglement entropy, and correlation functions. Our work offers a new platform to explore emergent non-thermal dynamics in state-of-the-art quantum simulators with gauge symmetries.
\end{abstract}

\maketitle

\section{Introduction}
Quantum thermalization is prevalent in quantum many-body physics. It refers to the phenomenon that the long-time dynamical behavior of a closed quantum system can be described by a thermal ensemble characterized by a few parameters such as temperature and chemical potential, accompanied by the loss of the local information of the initial state \cite{Nandkishore2015,Mori2018,Ueda2020,Abanin2019}. Two important classes of exceptions are known to severely break quantum thermalization, one is the quantum integrable systems with the number of conserved quantities being equal to the degree of freedoms \cite{Sutherland2004,Rigol2007}, and the other is the  disordered systems which support the many-body localization (MBL) \cite{Fleishman1980,Znidaric2008,Pal2010}. A strongly disordered system typically carries a set of local integrals of motion which localizes the excitations and freezes the transport \cite{Serbyn2013,Huse2014,Imbrie2016,Imbrie2017}, allowing the local information of the initial states to survive for a long time without being erased. These features also underlie several potential applications of MBL states in quantum information processing. Over the past decade, MBL has been extensively studied in various contexts of physical systems, including cold atoms in optical lattices \cite{Polkovnikov2011,Schreiber2015,Choi2016}, trapped ions \cite{Smith2016}, nuclear magnetic resonance \cite{Wei2018}, superconducting circuits \cite{Orell2019,Neill2018} and so on.

In recent years, another interesting mechanism of non-thermalization has been found in lattice gauge models without disorders, namely the disorder-free MBL \cite{Smith2017,Brenes2018,Papaefstathiou2020,Karpov2021,Halimeh1,Lang2022,Halimeh2}. In these systems, the quantum dynamics are constrained by local gauge symmetries, causing a portion of the system effectively to experience a disorder under the gauge-sector average. Particularly for the lattice quantum electrodynamics (QED) [also called the lattice Schwinger model (LSM)] with gauge fields being realized by high spins ($S=1$) \cite{Brenes2018}, fermions (matter fields) fail to thermalize when relaxed from a clean N\'{e}el state. This MBL results from the combined effect of the U(1) gauge symmetry (the Gauss's Law) and the electric field energy $E^2$ in the Hamiltonian. However, in the LSM with gauge fields being spin-1/2, a system that has recently been realized in two cold-atom simulators \cite{Yang2020,Mil2020,Zhou2022}, this disorder-free MBL induced by the electric field energy would not occur due to the vanishing of the $E^2$ term.

In this paper, we show that, contrary to what has been described above, the gauge fields of the spin-1/2 LSM can in fact also exhibit non-thermal dynamics, such as disorder-free MBL and prethermalization, as long as the system carries a four-fermion interaction term. Including such a fermion interaction into the Schwinger model was motivated by a recent proposal on realizing the synthetic U(1) gauge field using spin-1 bosons \cite{Gao2022}, in which the four-fermion interaction naturally arises from the intrinsic interactions of spinor cold atoms.
With the help of Gauss's Law, the fermion interaction can be transformed away which gives rise to a type of effective disorder for the gauge particles, rendering the latter to exhibit MBL dynamics. We carry out detailed numerical simulations on such quantities as local observables, bipartite entanglement entropy and correlation functions, in which the dynamical features of thermalization breaking can be clearly demonstrated.

The rest of this paper is organized as follows: In Sec.~\ref{confinedMBL}, we introduce the U(1) lattice Schwinger model and briefly review the mechanism of the disorder-free MBL. In Sec.~\ref{deconfinedMBL}, we present our scheme for breaking the thermalization by four-fermion interactions in a spin-1/2 LSM. In Sec.~\ref{results}, we go into detail about our numerical results. A brief conclusion can be found in Sec.~\ref{Summary}.

\section{Dynamical MBL in the High-Spin LSM} \label{confinedMBL}
\begin{figure}[t]
	\includegraphics[width=0.46\textwidth]{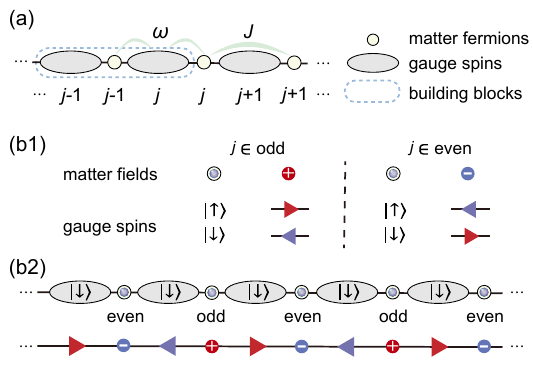}
	\caption{(a) Schematic of the LSM [Eq.~(\ref{LSM})] and that with four-fermion interaction [Eq.~(\ref{ILSM})]. The round circles denote the matter fields, and ovals denote the gauge fields. The blue dashed line labels a building block consisting of two neighboring gauge fields and one matter field in the middle. $\omega$ indicates the coupling between matter and the gauge fields; $J$ denotes the fermion interaction between two nearest-neighboring matter fields. (b) LSM and its QED picture in the framework composed of particles and anti-particles, i.e., $\tilde{H}$ in Eq.~(\ref{Hdd}). (b1) The correspondence between matter (gauge) fields in the left column and the charges (electric fields) in the right column. Specifically, a matter field occupation on odd (even) sites corresponds to the generation of a positron (electron) with a positive (negative) electric charge in the vacuum. An up-polarized gauge spin on odd (even) sites denotes the right-moving (left-moving) electric fields. The electric directions are inverted for a down-polarized gauge spin. (b2) A detailed example of a state (upper row), with all the matter sites being occupied and all the gauge spins being down-polarized, and its QED picture (bottom row). }
	\label{Fig1}
\end{figure}

Before fully engaging in our scheme, we briefly review the disorder-free MBL in the LSM with high-spin gauge fields. The continuous Schwinger model refers to the (1+1)D QED theory with U(1) gauge invariance, depicting the interactions between electrons (matter fermions) and photons (gauge bosons). It is also widely used as a toy model to study various phenomena in quantum chromodynamics, such as quark confinement and chiral symmetry breaking \cite{Schwinger1951,Schwinger1953,Coleman1976}. The lattice Hamiltonian  of the Schwinger model can be obtained by following the discretization convention provided by Kogut and Susskind \cite{Kogut1975}, which is formalized as (setting $\hbar=1$)
\begin{equation}
\begin{aligned}
H_\text{LSM} =& -\omega\sum_{j} \left(\psi_{j-1}^\dagger U_{j} \psi_{j} +  \text{h.c.}\right) \\
&+ m \sum_{j} (-1)^j  \psi_j^\dagger \psi_j + \frac{g^2}{2}\sum_j E_{j}^2,
\end{aligned}
\label{LSM}
\end{equation}
where ${\psi}_j^\dagger$ (${\psi}_j$) indicate the local matter fields of charged fermions, $j\in \mathbb{Z}^+= \{1,2,...,L\}$ with $L$ being the length of the chain; $U_{j}$ and $E_{j}$ satisfy the $\mathfrak{su}$(2)-like algebra $[E_j,U_k]=\delta_{j,k}U_j$, and denote respectively the parallel transporter and the electric field of the gauge fields living on the link between two neighboring matter sites ${\psi}_{j-1}$ and ${\psi}_j$, as is schematically shown in Fig.~\ref{Fig1}(a). In $H_\text{LSM}$, the first term describes the coupling between matter and gauge fields with coupling strength $\omega$, and the second term is the staggered mass referring to the opposite mass experienced by fermions seated on odd and even sites. The occurrence of negative mass is somewhat strange. However, as we will show later in Sec.~\ref{deconfinedMBL}, by introducing the antiparticles, the mass term will have a clearer picture --- two neighboring fermions respectively correspond to the electron and the positron carrying opposite charges but the same mass. The last term of $H_\text{LSM}$ indicates the energy of the gauge field with $g^2>0$ the coupling constant, which is purely composed of the electric energy $E^2$. This is a property of (1+1)D where the magnetic field is absent since the curl of the vector potential field is forbidden in one-dimensional space. In quantization, the electric states can only take integer values up to a shift, i.e., $E_j=\mathbb{Z}-\theta/2\pi$, where $\theta\in [0,2\pi)$ is the topological angle indicating a background electric field \cite{Coleman1976, Cheng2022, Halimeh2022}.

The LSM [Eq.~(\ref{LSM})]  carries a local gauge symmetry $[G_j,H_\text{LSM}]=0$, with
\begin{equation}
 G_j=  \psi_j^\dag \psi_j-(E_{j+1}- E_{j}) + \frac{1}{2}[(-1)^j-1]\,,
\label{Gf}
\end{equation}
being the Gauss operator defined in a building block consisting of two gauge fields $\{E_{j},E_{j+1}\}$ and one matter field $\psi_j$ in the middle [see Fig.~\ref{Fig1}(a)]. The static charge $q_j$ is defined as the quantum number of $G_j$, which is apparently a good quantum number. Up to some constants, $q_j$ locally characterizes the difference between the net electric flux $E_{j+1}- E_{j}$ and the fermionic charge $\psi_j^\dag \psi_j$, which is a direct manifestation of the Gauss's Law. The local gauge symmetry divides the entire Hilbert space into different gauge sectors, with each gauge sector labeled by a set of static charge numbers $\mathbf{q}=\{q_1,q_2,...,q_L\}$.

To quantum simulate the LSM in experiments \cite{Yang2020,Mil2020,Zhou2022}, it is common to realize the electric fields by spin-$S$ spinors, i.e., $U^{(\dag)}_j\rightarrow S_j^{\pm}$ and $E_j\rightarrow S_j^{z}$, which is also called the quantum link model \cite{Chandrasekharan1997,Wiese2013} or spin-gauge model \cite{Zohar2012,Zohar2016}. This means selecting a finite-dimensional representation for the $\mathfrak{su}$(2) gauge fields and truncating $E_j$ within the range $[-S,S]$. For high-spin LSM, the matter fields can exhibit the disorder-free MBL as the gauge fields are integrated out. The electric energy term $E_{j}^2$ in $H_\text{LSM}$ is responsible for this phenomenon. Let us take $S=1$ as an example \cite{Brenes2018}. In each gauge sector, the gauge field can be expressed by the static charge $q_j$ and the matter-field occupation $\psi^\dagger_j \psi_j$ using Gauss's Law [Eq.~(\ref{Gf})]. Consequently, the electric energy can be re-expressed completely in terms of the matter fermions, and contains a term $H_\text{d}=\sum_j q'_j \psi^\dagger_j \psi_j$, with $q'_j(\mathbf{q})$ being a function depending on the gauge-sector number $\mathbf{q}$. Therefore, if the initial state $\left\vert\Psi_0\right\rangle$ spans over a large number of random gauge sectors, $H_\text{d}$ effectively acts like a disorder potential, rendering the post-quench dynamics to break the thermalization. However, this mechanism would no longer work in the spin-1/2 LSM, since in this case, $E_j^2= (\sigma_j^{z})^2/4=1/4$ (with $\sigma_j^z$ being the spin-1/2 Pauli matrix) is simply a constant that can be neglected. Hence, the electric-energy-based approach in inducing the disorder-free MBL will not be applicable.

We additionally would like to mention that in the gauge sector $\mathbf{q}=\mathbf{0}$, the $E^2$ term is also closely related to the charge confinement of QED \cite{Banerjee2012,Zohar2016,Coleman1976}. Particularly, the high-spin LSM ($\theta\neq \pi$) is confined.
Separating two fermions with opposite charges would lead to an electric string between them, and hence the total electric energy in the Hamiltonian would be linearly proportional to the length of the string, i.e., $\propto |i-j| g^2 S^2$. To lower the electric energy, an additional pair of fermionic charges will emerge to screen the electric string, which is known as the string breaking \cite{Banerjee2012,Zohar2016,Magnifico2020}. In contrast, the LSM with $S=1/2$ ($\theta = \pi$) is deconfined due to the absence of the electric energy in $H_\text{LSM}$. In this case, the state of a local electric field should be either $1/2$ or $-1/2$, causing the total electric energy to be a constant $L g^2 S^2/2$ independent of the distribution of fermions. Note that not all gauge sectors are equivalent for the LSM with a finite $S$, and hence the relationship between the confinement and the disorder-free localization remains an open question worthy of further study. However, this question goes beyond the scope of our current paper.

%In the {\color{red}high-spin LSM}, the matter fields can exhibit the disorder-free MBL as the gauge fields are integrated out. The $E_{j}^2$ term plays a significant role in such a process. Let us take $S=1$ as an example \cite{Brenes2018}. In each gauge sector, the gauge field can be expressed by the static charge $q_j$ and the matter-field occupation $\psi^\dagger_j \psi_j$ using Gauss's Law [Eq.~(\ref{Gf})]. Consequently, the electric energy can be re-expressed completely in terms of the matter fermions, and contains a term $H_\text{d}=\sum_j q'_j \psi^\dagger_j \psi_j$, with $q'_j(\mathbf{q})$ being a function depending on the gauge-sector number $\mathbf{q}$. Therefore, if the initial state $\left\vert\Psi_0\right\rangle$ spans over a large number of random gauge sectors, $H_\text{d}$ effectively acts like a disorder after sector average, rendering the post-quench dynamics to break the thermalization. However, this mechanism would no longer work in the {\color{red}spin-1/2 LSM} due to the absence of the confined electric energy in the Hamiltonian, as we discussed above. As of now, state-of-the-art experimental techniques can only synthesize the spin-1/2 LSM \cite{Yang2020,Mil2020}, prompting us to think about the possible ways to break the thermalization in such a system.

\section{Dynamical MBL in the Spin-1/2 LSM} \label{deconfinedMBL}
Considering the fact that current state-of-the-art experimental techniques can only realize the spin-1/2 LSM \cite{Yang2020,Mil2020,Zhou2022}, it is highly desirable to investigate possible ways to break the thermalization in such a system.
Here, we formally discuss our scheme. The basic idea is to introduce a four-fermion interaction term into $H_\text{LSM}$ such that the total Hamiltonian now reads,
\begin{equation}
\begin{aligned}
H =& -\omega\sum_{j} \left(\psi_{j-1}^\dagger S_{j}^+ \psi_{j} +  \text{h.c.}\right) \\
&+ m \sum_{j} (-1)^j  \psi_j^\dagger \psi_j +  J\sum_{j} {\psi}_{j-1}^\dagger {\psi}_{j-1} {\psi}_{j}^\dagger {\psi}_{j},
\end{aligned}
\label{ILSM}
\end{equation}
with $J$ characterizing the interaction strength. The addition of this interaction term does not affect the local gauge symmetry as $[G_j,H]=0$ still holds. Our goal is to show that, by integrating the matter field out, the gauge field experiences an effective disorder. Our approach thus is in contrast to the scheme shown in Ref.~\cite{Brenes2018} where the MBL is realized on matter fields. Introducing the fermion interaction term into the LSM was proposed in Ref.~\cite{Gao2022}, in which the equilibrium-state phase diagram and quench dynamics were studied under a fixed gauge sector $\mathbf{q}=\mathbf{0}$. In the current work, we find that the fermion interaction is capable of inducing non-thermal dynamics when different gauge sectors are mixed.

We explicitly introduce the anti-particles by taking the particle-hole transformation on the odd sites, i.e., $ \psi_{j\in\text{odd}}\rightarrow\psi_{j\in\text{odd}}^\dagger$, and making a similar transformation on the gauge fields $ S_{j\in\text{odd}}^+\rightarrow- S_{j\in\text{odd}}^-,  S_{j\in\text{odd}}^z\rightarrow- S_{j\in\text{odd}}^z$, which transforms Hamiltonian (\ref{ILSM}) into a new form:
\begin{equation}
\begin{aligned}
\tilde{H}=&-\omega\sum_{j}\left( \psi_{j-1} S _{j}^{+} \psi_{j}+\text{h.c.}\right) \\
&+m^{\prime}\sum_{j}
 \psi_j^\dag \psi_j-J\sum_{j} \psi_{j-1}^\dag \psi_{j-1} \psi_j^\dag \psi_j ,
\end{aligned}
\label{Hdd}
\end{equation}
 with $m'=m+J$. Correspondingly, we have the Gauss operator
\begin{equation}
\tilde G_j= \psi_j^\dag \psi_j+{S}_{j}^z+{S}_{j+1}^z,\\
\label{Gq}
\end{equation}
with $[\tilde G_j,\tilde H]=0$, and $\mathbf{\tilde q}=\{\tilde q_1,\tilde q_2,...,\tilde q_L\}$ labels the gauge sectors with $\tilde q_j$ being the quantum number of $\tilde G_j$. Apparently, $\tilde{H}$ is translationally invariant with all fermions featuring the same mass $m'$, as mentioned before. $\tilde{H}$ provides a clear analog of the LSM in QED [see Fig.~\ref{Fig1}(b1)]: The occupation of the odd and even matter sites respectively denote the positron and electron with equal mass $m'$; for gauge spins at even sites, states $\left|\uparrow\right\rangle$ and $\left|\downarrow\right\rangle$ respectively correspond to the left- and right-pointing electric fields; whereas for gauge spins at odd sites, the directions of electric fields are reversed. In Fig.~\ref{Fig1}(b2), we show a concrete example of state (upper row) with all the matter sites being occupied and its QED analog (bottom row), in which the distributions of charges and electric fields are clearly illustrated. In this picture, the matter-gauge interaction ($\omega$ term in $\tilde H$) indicates the process that a pair of electron and positron merge together simultaneously generating gauge photons. Photon generation in the context of $S=1/2$ corresponds to the spin flip of gauge spins. Also within this picture, Gauss's Law with $\tilde G_j$ indicates that the total excitation within a building block is conserved, including the electron (positron) and gauge spins.

Since matter fields and gauge spins are mutually related to each other by Gauss's Law [Eq.~(\ref{Gq})], we are in principle allowed to eliminate the matter fields and write down an effective model purely in terms of the gauge spins. Eliminating the matter fields is straightforward for the last two terms of $\tilde H$. To be specific, given a certain gauge sector $\mathbf{\tilde q}$, substituting Eq.~(\ref{Gq}) into Eq.~(\ref{Hdd}) leads to $-2m'\sum_j S_j^z -J\sum_j[\tilde q_{j-1}-({S}_{j-1}^z+{S}_{j}^z)][\tilde q_j-({S}_{j}^z+{S}_{j+1}^z)]$. The $m'$-term is free of disorder, and thus is irrelevant to the MBL dynamics. In the following discussion, we thus focus on the case of $m'=0$. In contrast, the $J$-term, arising from the fermion interaction, is gauge-sector relevant. Rewriting the $J$-term in terms of gauge spins yields
\begin{equation}
-J\sum_{j}(2 S_{j}^z S_{j+1}^z+ S_{j-1}^z S_{j+1}^z -\tilde{q}_j' S_{j}^z),
\label{Hdis}
\end{equation}
with $\tilde{q}_j'=\tilde{q}_{j-2}+\tilde{q}_{j-1}+\tilde{q}_{j}+\tilde{q}_{j+1}$. It indicates that, in addition to the homogeneous interactions ($S_{j}^z S_{j+1}^z$ and $S_{j-1}^z S_{j+1}^z$), the gauge field additionally experiences a local potential $-\tilde{q}_j' S_{j}^z$ whose strength depends on the gauge sector $\mathbf{\tilde q}$. Therefore, if the initial state mixes various random gauge sectors, the gauge spins would experience an effective disorder under sector average. This term therefore plays a central role in our scheme in inducing the anomalous non-thermal dynamics, as will be presented in Sec.~\ref{results}.

\begin{table}[]
\begin{tabular}{|c|c|c|c|c|}
\hline
$\tilde{q}_j$                                                                                                   & -1                                  & 0                                                                                                                                                     & 1                                                                                                                                                   & 2                                 \\ \hline
\begin{tabular}[c]{@{}c@{}} configurations\\ $\left|S_j^z,n_j,S_{j+1}^z\right\rangle$\end{tabular} & $\left|\downarrow,0,\downarrow\right\rangle$ & \begin{tabular}[c]{@{}c@{}}$\left|\uparrow,0,\downarrow\right\rangle$\\ $\left|\downarrow,0,\uparrow\right\rangle$\\ $\left|\downarrow,1,\downarrow\right\rangle$\end{tabular} & \begin{tabular}[c]{@{}c@{}}$\left|\uparrow,1,\downarrow\right\rangle$\\ $\left|\downarrow,1,\uparrow\right\rangle$\\ $\left|\uparrow,0,\uparrow\right\rangle$\end{tabular} & $\left|\uparrow,1,\uparrow\right\rangle$ \\ \hline
\end{tabular}
\caption{\label{Tab1} Allowed configurations $\left|S_j^z,n_j,S_{j+1}^z\right\rangle$ in the $j$-th building block, with $\tilde{q}_j$ being the quantum number of $\tilde{G}_j$.}
\end{table}

Within a building block as defined in Fig.~\ref{Fig1}(a), $\tilde q_j$ is allowed to take four integer values, i.e., $\tilde q_j\in\{-1,0,1,2\}$. By respectively choosing the Fock basis $|n_j = 0,1\rangle$ and the spin basis $|S_i^z=\uparrow,\downarrow\rangle$ for the matter and gauge fields, the correspondence between $\tilde q_j$ and the allowed configurations is listed in the Tab.~\ref{Tab1}. It can be observed that $\tilde q_j=\{0,1\}$ each possesses three distinct configurations, whereas $\tilde q_j=\{-1,2\}$ each possesses only one unique configuration. We thus consider an initial state
\begin{equation}
|\Psi_0\rangle= \left( \frac{|0\rangle+ |1\rangle}{\sqrt{2}}\right)^{\otimes L}|\downarrow,\uparrow,\downarrow,...\rangle,
\label{Psi0}
\end{equation}
which is a product state, with the matter fields being an equal superposition of states $|0\rangle$ and $|1\rangle$, and the gauge fields being simply an antiferromagnetic N\'{e}el state. In each building block, the state $|\Psi_0\rangle$ completely lies in $\tilde q_j=\{0,1\}$ with equal probability $1/2$. Hence, for a chain with length $L$, there are totally $2^L$ gauge sectors involved. Most of these gauge sectors are with a random $\mathbf{\tilde q}$, e.g., $\mathbf{\tilde q}=\{1,0,0,1,0,1,\cdots\}$. There are indeed some exceptions. For example, $\mathbf{\tilde q}=\mathbf{0}$ and $\mathbf{\tilde q}=\mathbf{1}$ are completely ordered. However, the portion of them is always exponentially small, and hence they would not dominate the dynamics for a large $L$.

Eliminating the matter fields from the first term of $\tilde{H}$ is not as straightforward as from the last two terms. Up to now, no simple way exists to eliminate the matter fields for a general random $\mathbf{\tilde q}$. However, as will be shown by numerics below, the $\omega$ term alone in Eq.~(\ref{Hdd}) is unable to prevent thermalization, manifested by the phenomenon that the local gauge spins of $|\Psi_0\rangle$ quickly relax to thermal equilibrium. Therefore, $|\Psi_0\rangle$ serves as an important reference state for the discussion of the thermalization breaking induced by the fermion interaction $J$. It may also be worthwhile to mention that, in the completely ordered gauge sectors ($\mathbf{\tilde q}=\mathbf{0}$ or $\mathbf{\tilde q}=\mathbf{1}$), matter-field elimination can be accomplished by mapping the system into a Rydberg chain \cite{Surace2020, Gao2022, Cheng2022, Bernien2017}. The resulting term is a PXP Hamiltonian which is known to possess a set of quantum many-body scar states weakly breaking the eigenstate thermalization hypothesis \cite{Turner2018, Serbyn2021}. In spite of this, the mapping cannot be simply generalized to a general $\mathbf{\tilde q}$. Since the weight of the ordered sectors is sufficiently small as mentioned above, we will not discuss this any further in this paper.

\section{Numerical Results} \label{results}
In practical simulations, it is convenient for us to additionally map the fermions of Eq.~(\ref{Hdd}) into Pauli spins using the Jordan-Wigner transformation:
\[ \psi^{\dagger}_j=s^+_j \, \prod_{l=1}^{j-1}(2n_l-1)\,,\;\;\;\psi_j=s^-_j\,\prod_{l=1}^{j-1}(2n_l-1)\,,\] with $n_l=s^+_ls^-_l$.
Under this mapping, $\tilde H$ can be written into an interacting spin chain Hamiltonian
\begin{equation}
H_s=-\sum_{j}\left[\omega\left( s^-_{j-1} S _{j}^{+} s^-_j+\text{h.c.}\right)+J s^z_{j-1}s^z_{j}+J s^z_{j}\right],
\label{Hs}
\end{equation}
in which the gauge spins and the matter spins are denoted by capital $S_j$ and lowercase $s_j$, respectively. In correspondence, the initial state has the form
\begin{equation}
|\Psi_0\rangle=\left(\frac{\left|\Uparrow\right\rangle+ \left|\Downarrow\right\rangle}{\sqrt{2}}\right)^{\otimes L}|\downarrow,\uparrow,\downarrow,...\rangle,
\label{Psi0}
\end{equation}
with $\left|\Uparrow\right\rangle$ and $\left|\Downarrow\right\rangle$ denoting the eigenstates of matter spins $s^z$.
We simulate the dynamics $|\Psi (t)\rangle = e^{-i H_s t} |\Psi_0\rangle$ via exact diagonalization of the Hamiltonian $H_s$. By utilizing the (discrete) translational symmetry of $H_s$ and $|\Psi_0\rangle$ \cite{Sandvik2010,Weinberg2017}, we are able to deal with a system of size up to $L=14$ (i.e., 14 matter spins plus 14 gauge spins) on a medium-size workstation.

We first look at the dynamics of local polarization of gauge spins, i.e., $\langle S_j^z(t) \rangle$. Generally for a many-body system under thermalization \cite{Yao2016,Schiulaz2015,Enss2017}, after a sufficiently long time of evolution, all the local information of the initial state would be erased and the system would behave like a thermal state characterized by density matrix $\rho_\text{th}$. Namely, the local observable $\langle S_j^z(t) \rangle$ would approach the thermal equilibrium, i.e.,
\begin{equation}
\lim_{t\rightarrow \infty} \langle S_j^z(t) \rangle\approx\langle S_j^z \rangle_\text{th}=\mathrm{Tr}(\rho_\text{th} S_j^z)\,,
\end{equation}
with
\begin{equation}
\rho_\text{th}=\frac{e^{-\beta H_s}}{\text{Tr}(e^{-\beta  H_s})}
\label{rhoth}
\end{equation}
being the density matrix of the Gibbs ensemble, with $\beta$ the effective inverse temperature determined by the initial state via $\mathcal{E}={\langle} \Psi_0 | H_s | \Psi_0 \rangle=\text{Tr}(\rho_\text{th} H_s)$. In contrast, for systems breaking the thermalization, such as the MBL, the local equilibration $\lim_{t\rightarrow \infty} \langle S_j^z(t) \rangle$ would deviate from the thermal value $\langle S_j^z \rangle_\text{th}$. Our numerics show that, for arbitrary $J$, the thermal state $\rho_\text{th}$ associated with our initial state $|\Psi_0\rangle$ [Eq.~(\ref{Psi0})] is always an infinite-temperature thermal state, i.e., $\rho_\text{th} \propto \mathbb{I}$, such that $\langle S_j^z \rangle_\text{th}=0$. This can be understood in the following way. Since $|\Psi_0\rangle$ is a product state with each matter spin being $(\left|\Uparrow\right\rangle + \left|\Downarrow\right\rangle)/\sqrt{2}$ and each gauge spin being either $\left|\uparrow\right\rangle$ or $\left|\downarrow\right\rangle$, it thus has zero energy expectation $\mathcal{E}={\left\langle\Psi_0\right\vert H_s\left\vert\Psi_0\right\rangle}=0$. On the other hand, $H_s$ is traceless such that the average of all the eigen energies is also equal to zero. These two facts indicate that $\mathcal{E} = \mathrm{Tr}(\rho_\text{th} H_s)=0$ should occur at $\beta = 0$, namely at the infinite temperature. The infinite-temperature state should have vanishing expectation values for all the traceless operators, and hence the deviation of the long-time dynamics of local traceless operators from zero conveniently measures the degree of thermalization breaking.

\begin{figure}[t]
	\includegraphics[width=0.48\textwidth]{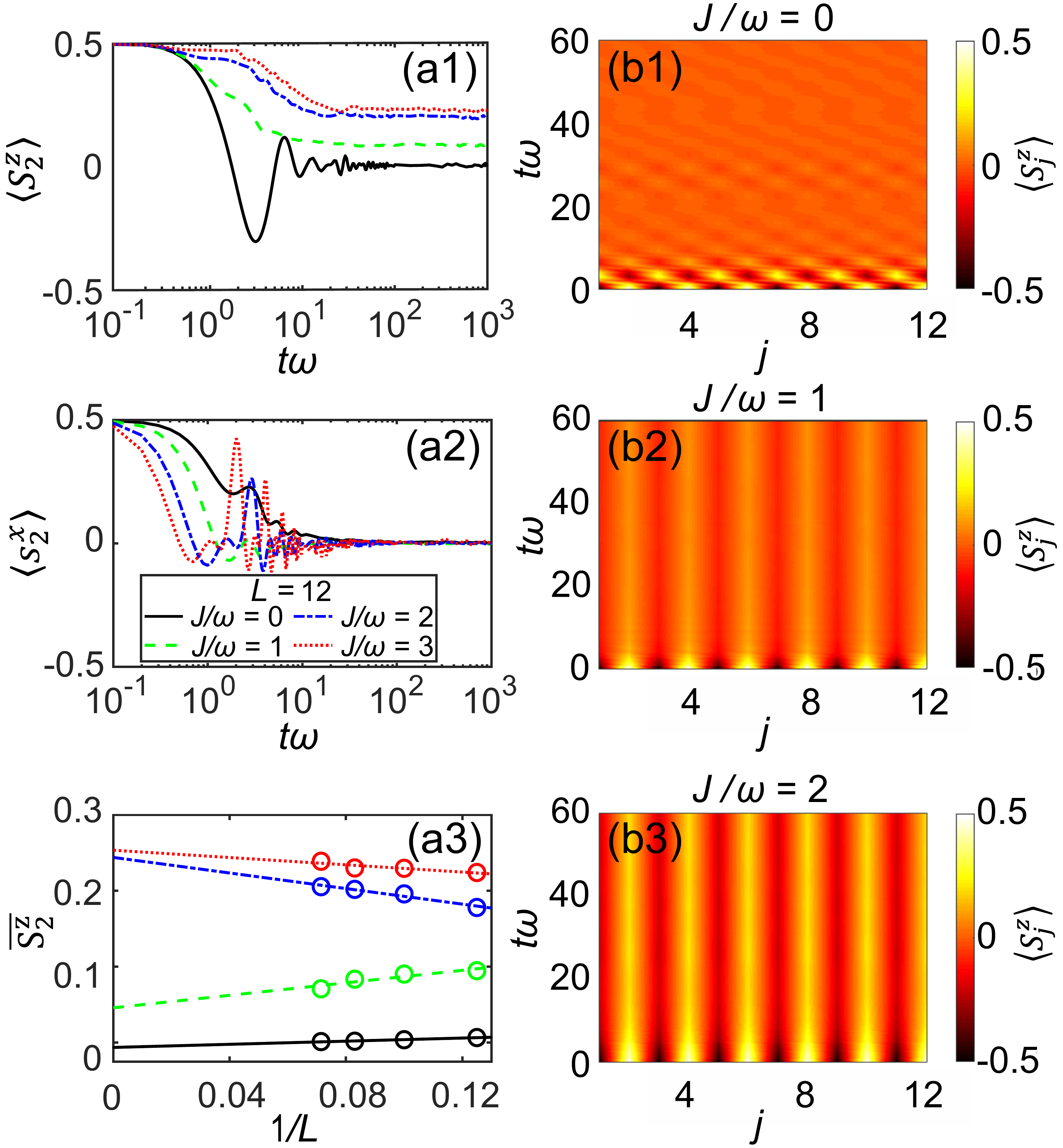}
	\caption{(a) Time evolution of a local gauge spin $\left\langle S^z_{j=2}\right\rangle$ (a1) and a local matter spin $\left\langle s^x_{j=2}\right\rangle$ (a2) on $\log(t)$, with solid, dashed, dot-dashed and dotted lines corresponding to the cases of $J=0$, $\omega$, $2\omega$ and $3\omega$, respectively. The (a3) indicates the averaged long-time polarization $\overline{S^z_2}$ [Eq.~(\ref{ASz})] versus system size, where markers are numerical data at $L=\{8,10,12,14\}$ and lines denote the linear fitting for the data. (b) Dynamics of local gauge polarizations $\left\langle S^z_{j}\right\rangle$ on each site $j$, where (b1), (b2) and (b3) denote the cases of $J=0$, $\omega$ and $2\omega$, respectively. Except for (a3), all other panels are calculated at $L=12$.}
	\label{Fig2}
\end{figure}

In Fig.~\ref{Fig2}(a1), we plot the polarization of a local gauge field $\langle S_{j=2}^z(t) \rangle$ for various matter-field interaction $J$, with the solid, dashed, dot-dashed, and dotted lines denoting $J=0$, $\omega$, $2\omega$, and $3\omega$, respectively. One can observe that, in the absence of fermion interaction ($J=0$), the local polarization rapidly decays from $1/2$ to $\langle S^z_j \rangle_\text{th}=0$, as a manifestation of quantum thermalization. However, for the cases of $J\neq0$, the long-time behaviors $\langle S_{j=2}^z(t) \rangle$ apparently deviate from zero. With an increase in $J$, the deviation would become larger. These behaviors are consistent with our previous discussion that increasing $J$ leads to an increase in the disorder strength, which results in more severe destruction of quantum thermalization.

By contrast, the matter field does not exhibit thermalization breaking. This is show in Fig.~\ref{Fig2}(a2), where we plot the dynamics of $\langle s_{j=2}^x(t) \rangle$ for various values of $J$. As one can see, in the long-time limit, $\langle s_{j=2}^x(t) \rangle$ all converges to the thermal equilibrium value which is also zero, regardless of the values of $J$. This is understandable since the matter fields do not experience the disorder potential, which thus differs from the gauge fields.

Furthermore, to characterize the dependence of local polarization on the system size $L$, we perform a system-size analysis of the averaged polarization $\overline{S^z_2}$ and show the result in the Fig.~\ref{Fig2}(a3), where
\begin{equation}
\overline{S^z_2} = T^{-1}\int_{t_0}^{t_0+T} {dt \langle S^z_2(t) \rangle}
\label{ASz}
\end{equation}
with $t_0=50\omega^{-1}$ and $T=300\omega^{-1}$ chosen to be sufficiently large to ensure that $\overline{S^z_2}$ can capture the averaged long-time feature of the local gauge spin. It can be observed that, for large $J$ ($J=2\omega$ and $3\omega$), the local polarization increases slowly with system size, which indicates that the system is not ergodic in the thermodynamic limit.

The thermalization process is generally accompanied by the information loss of the initial state, which can be observed in the dynamics of gauge spins as shown in Fig.~\ref{Fig2}(b1)-(b3). We show the dynamics of $\langle S_{j}^z(t) \rangle$ for each site $j$, with panels (b1), (b2) and (b2) corresponding to $J=0$, $\omega$, and $2\omega$, respectively. At $t=0$, the staggered magnetization for the initial N\'{e}el state of gauge spins [Eq.~(\ref{Psi0})] is quite obvious. As time passes, the staggered magnetization structure vanishes for the case of $J=0$ [Fig.~\ref{Fig2}(b1)], indicating the information loss of the initial state. In contrast, for the non-thermal dynamics with $J=\omega$ and $2\omega$ [Fig.~\ref{Fig2}(b2), (b3)], the staggered magnetization structure persists after a long time of evolution. Moreover, the larger $J$ is, the more information about the initial state has remained.

\begin{figure}[t]
	\includegraphics[width=0.48\textwidth]{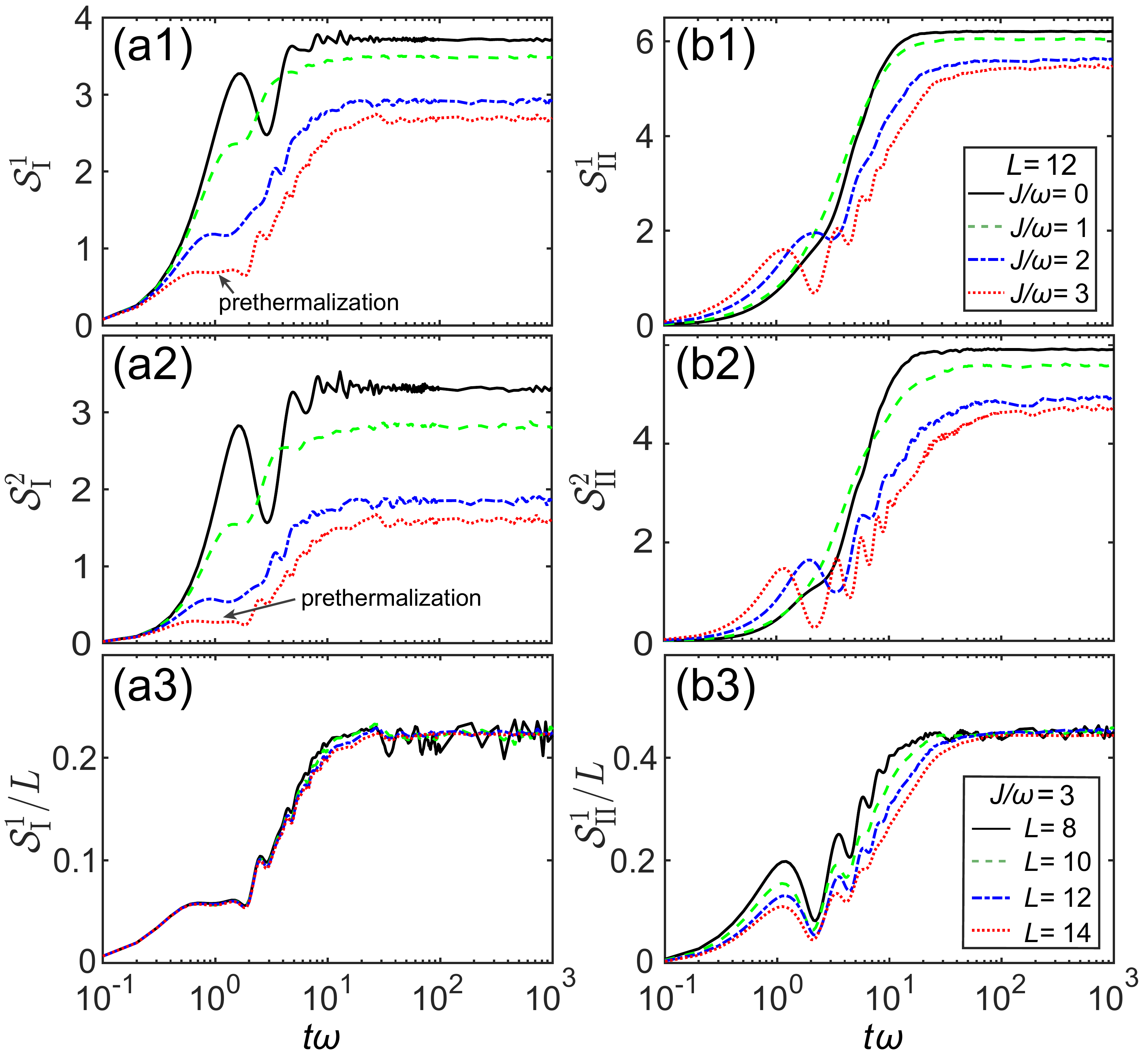}
	\caption{Bipartite entropy dynamics versus $\log(t)$, where the columns (a) and (b) respectively correspond to two ways (I and II) of partition of the system. Panels (a1) and (b1) correspond to the von Neumann entropy $S^1_\text{I}(t)$ and $S^1_\text{II}(t)$, where the solid, dashed, dot-dashed and dotted lines indicate the cases of $J=0$, $\omega$, $2\omega$ and $3\omega$, respectively. Panels (a2) and (b2) correspond to the 2nd-order R\'{e}nyi entropy $S^2_\text{I}(t)$ and $S^2_\text{II}(t)$. Panels (a3) and (b3) respectively show the dependence of $S^1_\text{I}(t)/L$ and $S^1_\text{II}(t)/L$ on various lattice sizes $L$ at a fixed $J=3\omega$, with solid, dashed, dot-dashed and dotted lines denoting the cases of $L=8$, $10$, $12$ and $14$. }
	\label{Fig3}
\end{figure}

To characterize the entropy growth in the system, we calculate the dynamics of R\'{e}nyi entropy
\begin{equation}
{\cal S}_\text{I,II}^\alpha(t)=\frac{1}{1-\alpha}\log\mathrm{Tr}\rho_\text{A}^\alpha(t),
\label{Renyi}
\end{equation}
where $\rho_\text{A}=\mathrm{Tr}_\text{B} \rho = \mathrm{Tr}_\text{B} | \Psi \rangle\langle\Psi|$ is the reduced density matrix of the subsystem A, and $\alpha$ is the order of R\'{e}nyi entropy. Particularly in the limit of $\alpha \rightarrow 1$, the R\'{e}nyi entropy reproduces the von Neumann entropy \cite{Lennert2013}, i.e., ${\cal S}^1_\text{I,II}=-\mathrm{Tr} (\rho_\text{A} \log \rho_\text{A})$. The subscription I and II indicate two different ways of partition the system: I) A consists of left half of gauge spins, while B the rest (i.e., right half of gauge spins and all matter spins); II) A consists of the left half of system including both gauge and matter spins, while B the right half of the system. In partition I, the boundary between the two subsystems is extensive, while in partition II, the boundary is not extensive since it is just a single site as the entire chain is cut into two halves directly from the middle.

%In the process of quantum thermalization, the R\'{e}nyi entropy would be linearly proportional to time, i.e., ${\cal S}^\alpha(t)\propto t$, whereas for the MBL dynamics, the R\'{e}nyi entropy slowly grows in a logarithmic way ${\cal S}^\alpha(t)\propto \log (t)$ \cite{Nandkishore2015,Mori2018}.

In the upper two rows of Fig.~\ref{Fig3}, we fix $L=12$ and show respectively the dependence of the von Neumann entropy $S^1_\text{I,II}$ and the 2nd-order R\'{e}nyi entropy ${\cal S}^2_\text{I,II}$ on $\log(t)$, where different line styles again indicate the cases of different fermion interaction $J$. Clearly, for a given partition (I or II), $\mathcal{S}^1$ and $\mathcal{S}^2$ exhibit similar behavior, allowing us to focus solely on the first row. $S^1_\text{I}$ and $S^1_\text{II}$ exhibit a similar long-time behavior after equilibration, i.e., the entropy saturates at a value ${\cal S}_\text{sat}$. ${\cal S}_\text{sat}$ decreases as $J$ increases, akin to the results observed in the conventional disorder-free MBL \cite{Brenes2018,Halimeh1,Lang2022}. However, in the short-time scale, $S^1_\text{I,II}$ exhibit some unconventional features. Particularly in Fig.~\ref{Fig3}(a1), for $J=0$, $S^1_\text{I}$ shows a smooth and rapid growth with speed faster than $\log(t)$; whereas for large $J$ (e.g., $J=3\omega$), $S^1_\text{I}$  first hits a small plateau $\mathcal{S}_\text{pre}$, and then increases approximately linearly in $\log(t)$ until saturation. The small plateau $\mathcal{S}_\text{pre}$ is called the prethermalization \cite{Berges2004,Nandkishore2015,Mori2018,Ueda2020} indicating the gauge spins exhibit an intermediate quasi-stationary state before being further thermalized. The prethermalization plateau becomes more and more obvious as $J$ grows. $S^1_\text{II}$ in Fig.~\ref{Fig3}(a2) is similar to $S^1_\text{I}$ in the short time, but there exists the difference which mainly lies in that $S^1_\text{II}$ oscillates during the prethermalization stage of $S^1_\text{I}$.

In Fig.~\ref{Fig3}(a3) and (b3), we fix $J=3\omega$ and show respectively the dependence of ${\cal S}^1_\text{I}(t)/L$ and ${\cal S}^1_\text{II}(t)/L$ on various system sizes $L$, with solid, dashed, dot-dashed and dotted lines correspond to the cases of $L=8$, $10$, $12$ and $14$, respectively. The long-time feature of the two figures is quite similar that all curves roughly collapse into a single curve, indicating extensive entropy saturation. On the other hand, the short-time behavior of Fig.~\ref{Fig3}(a3) is also extensive, whereas that of (b3) is non-extensive. The discrepancy can be attributed to the ways on how the system is partitioned. As we mentioned earlier, the boundary between the two subsystems is extensive (non-extensive) for partition I (II).

\begin{figure}[t]
	\includegraphics[width=0.49\textwidth]{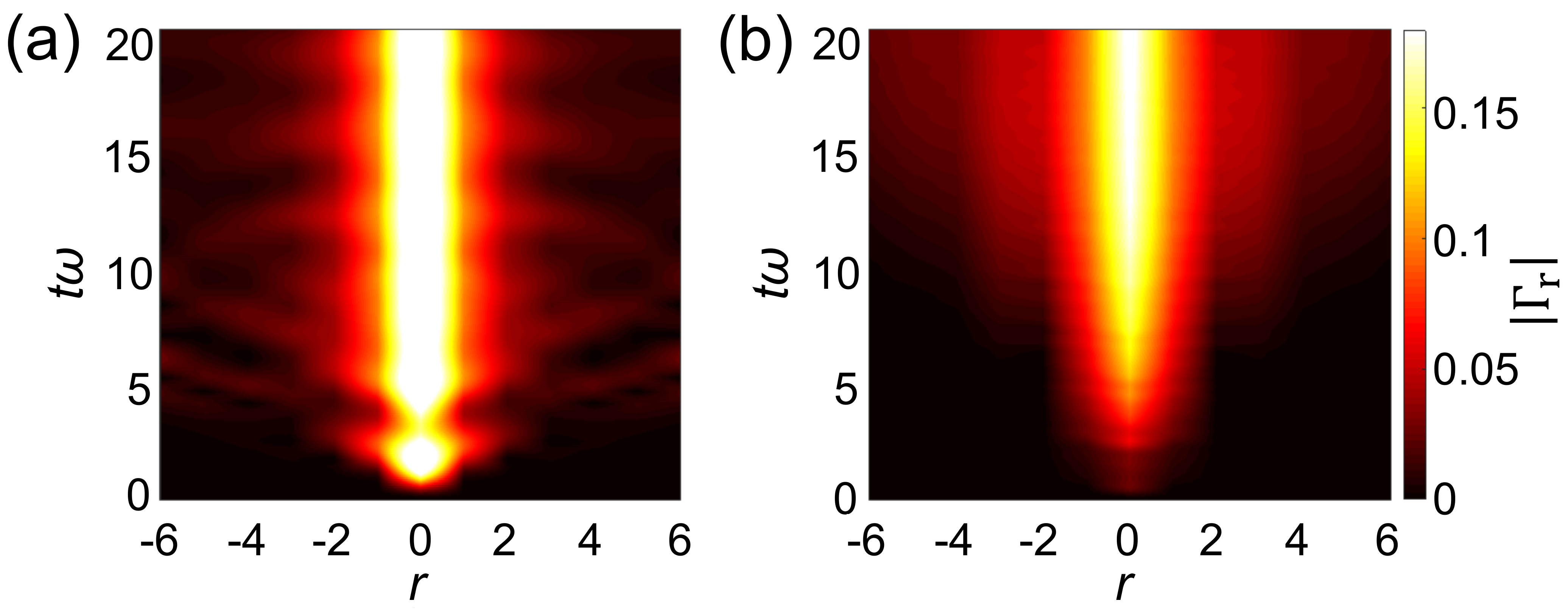}
	\caption{Dynamics of the connected correlation function $\Gamma_r$ of gauge spins, with $r$ being the distance between two spins. (a) The case of $J=0$. (b) The case of $J=3\omega$. In the calculation, we fix $L=14$.}
\label{Fig4}
\end{figure}
The magnitude of thermalization can be also reflected in the propagation of correlators.
In practice, we calculate the connected two-point correlation function of gauge spins:
\begin{equation}
\begin{aligned}
\Gamma_r(t) =\left\langle  S^z_j(t) S^z_{j+r}(t)\right\rangle-\left\langle  S^z_j (t)\right\rangle\left\langle  S^z_{j+r} (t)\right\rangle,
\end{aligned}
\label{sc}
\end{equation}
with $r$ denoting the relative distance. The results of the cases $J=0$ and $J=3\omega$ are shown in Figs.~\ref{Fig4}(a) and (b), respectively. One can observe that $\Gamma_r$ is zero at $t=0$ since the initial state $|\Psi_0\rangle$ is a product state and also an eigenstate of $S^z_j$. As $t$ increases, $\Gamma_r$ spreads out from the center to both sides.

One apparent feature is that the correlation propagation of $J=3\omega$ is much slower than that of $J=0$, which is consistent with our expectation on MBL \cite{Nandkishore2015}.
Generally, for a thermalizing system, correlation propagates ballistically forming a light cone $|r|\sim t$. In contrast, due to the exponential decay of interaction strength of localized dressed spins, the light core of the MBL is generally in the shape of $|r|\sim \log(t)$. The correlation boundary in Figs.~\ref{Fig4}(a) and (b) qualitatively capture the ballistic and $\log(t)$ light cones, respectively.

\section{Conclusion} \label{Summary}
We have shown that the four-fermion interaction term in the spin-1/2 lattice Schwinger model is responsible for the breaking of quantum thermalization. Under the gauge sector average, the gauge spins effectively experience a disorder after the matter degree of freedom is integrated out. This fermion-interaction-induced disorder underlies such non-thermal dynamics as many-body localization and entropy prethermalization when the system relaxes from an antiferromagnetic state. Our work promisingly facilitates the observation of disorder-free many-body localization in state-of-the-art cold-atom quantum simulators with U(1) gauge invariance.

\begin{acknowledgments}
L. C. acknowledges support from the NSF of China (Grants Nos. 12174236 and 12147215);
Y. Z. acknowledges support from the NSF of China (Grant No. 12074340) and the Science Foundation of Zhejiang Sci-Tech University (Grant No. 20062098-Y.);
H. P. acknowledges support from the US NSF and the Welch Foundation (Grant No. C-1669).
\end{acknowledgments}

\end{document}